\documentclass{ws-procs975x65}
\usepackage{subfigure}
\def\be{\begin{eqnarray}}
\def\ee{\end{eqnarray}}

\begin{document}

\title{QUANTUM COMPUTING AND \\THE ENTANGLEMENT FRONTIER}

\author{JOHN PRESKILL}

\address{Institute for Quantum Information and Matter\\
California Institute of Technology\\
Pasadena, CA 91125, USA}

\begin{abstract}
Quantum information science explores the frontier of highly complex quantum states, the ``entanglement frontier.'' This study is motivated by the observation (widely believed but unproven) that classical systems cannot simulate highly entangled quantum systems efficiently, and we hope to hasten the day when well controlled quantum systems can perform tasks surpassing what can be done in the classical world. One way to achieve such ``quantum supremacy'' would be to run an algorithm on a quantum computer which solves a problem with a super-polynomial speedup relative to classical computers, but there may be other ways that can be achieved sooner, such as simulating exotic quantum states of strongly correlated matter. To operate a large scale quantum computer reliably we will need to overcome the debilitating effects of decoherence, which might be done using ``standard'' quantum hardware protected by quantum error-correcting codes, or by exploiting the nonabelian quantum statistics of anyons realized in solid state systems, or by combining both methods. Only by challenging the entanglement frontier will we learn whether Nature provides extravagant resources far beyond what the classical world would allow. 
\begin{center}
{\em Rapporteur talk at the 25th Solvay Conference on Physics
\\
``The Theory of the Quantum World'' 
\\
Brussels, 19-22 October 2011}
\end{center}
\end{abstract}


\bodymatter

\section{Introduction: toward quantum supremacy}\label{sec:intro}
My assignment is to report on the current status of {\em quantum information science}, but I will not attempt to give a comprehensive survey of this rapidly growing field. In particular, I will not discuss recent experimental advances, which will be covered by other rapporteurs.

To convey the spirit driving the subject, I will focus on one Big Question:
\begin{quote}
\em{Can we control complex quantum systems and if we can, so what?}
\end{quote}
Quantum information science explores, not the frontier of short distances as in particle physics, or of long distances as in cosmology, but rather the frontier of highly complex quantum states, the {\em entanglement frontier}. I will address whether we can probe deeply into this frontier and what we might find or accomplish by doing so. This Big Question does not encompass everything of interest in quantum information science, but it gets to the heart of what makes the field compelling.

The quantum informationists are rebelling against a fundamental dualism we learned in school: 
\begin{quote}
\em{The macroscopic world is classical.\\
The microscopic world is quantum.}
\end{quote}
We fervently wish for controlled quantum systems that are large yet exhibit profoundly quantum behavior. The reason we find this quest irresistible can be stated succinctly:
\begin{quote}
\em{Classical systems cannot in general simulate quantum systems efficiently.}
\end{quote}
We cannot yet prove this claim, either mathematically or experimentally, but we have reason to believe it is true; arguably, it is one of the most interesting distinctions ever made between quantum and classical. It means that well controlled large quantum systems may ``surpass understanding,'' behaving in ways we find surprising and delightful. 

We therefore hope to hasten the onset of the era of {\em quantum supremacy}, when we will be able to perform tasks with controlled quantum systems going beyond what can be achieved with ordinary digital computers. To realize that dream, we must overcome the formidable enemy of {\em decoherence}, which makes typical large quantum systems behave classically. So another question looms over the subject: 
\begin{quote}
\em{Is controlling large-scale quantum systems merely {\bf really, really hard}, or is it {\bf ridiculously hard}?}
\end{quote}
In the former case we might succeed in building large-scale quantum computers after a few decades of very hard work. In the latter case we might not succeed for centuries, if ever.

This question is partly about engineering but it is about physics as well (and indeed the boundary between the two is not clearly defined). If quantum supremacy turns out to be unattainable, it may be due to physical laws yet to be discovered. In any case, the quest for large-scale quantum computing will push physics into a new regime never explored before. Who knows what we'll find?

\section{Quantum entanglement and the vastness of Hilbert space}
\begin{figure}
\begin{center}
\includegraphics[width=0.8\textwidth]{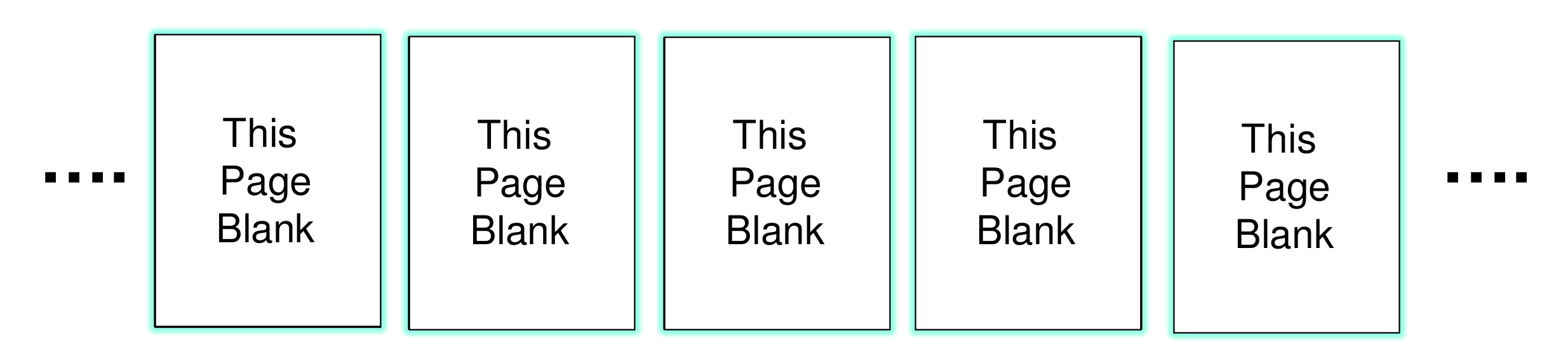}
\end{center}
\caption{\label{fig:quantum-book} For a typical quantum state with many parts, a measurement acting on just one part collects a negligible amount of information about the state.}
\end{figure}

At the core of quantum information science is entanglement, the characteristic correlations among the parts of a quantum system, which have no classical analog. We may imagine a quantum system with many parts, like a 100 page quantum book. If the book were classical, we could read 10 of the pages and learn about 10\% of the content of the book. But for a typical 100-page quantum book, if we read 10 pages we learn almost nothing about the content of the book; the information is not printed on the individual pages --- rather nearly all the information in the book is encoded in the correlations among the pages. (See Fig.~\ref{fig:quantum-book}.) These correlations are very complex, so that recording a complete classical description of the quantum state would require a classical book of astronomical size. 

Does Nature really indulge in such extravagant resources, and how can we verify it?


\begin{figure}
\centering
\mbox{\subfigure[]{\includegraphics[width=0.25\textwidth]{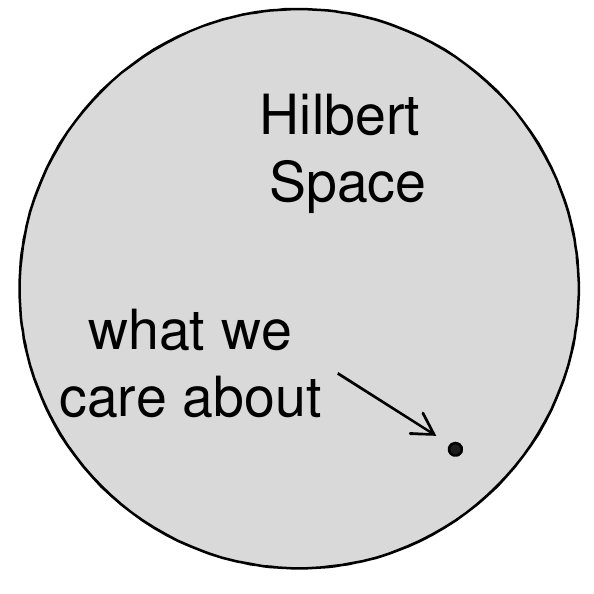}}
\hspace*{1in}
\subfigure[]{\includegraphics[width=.25\textwidth]{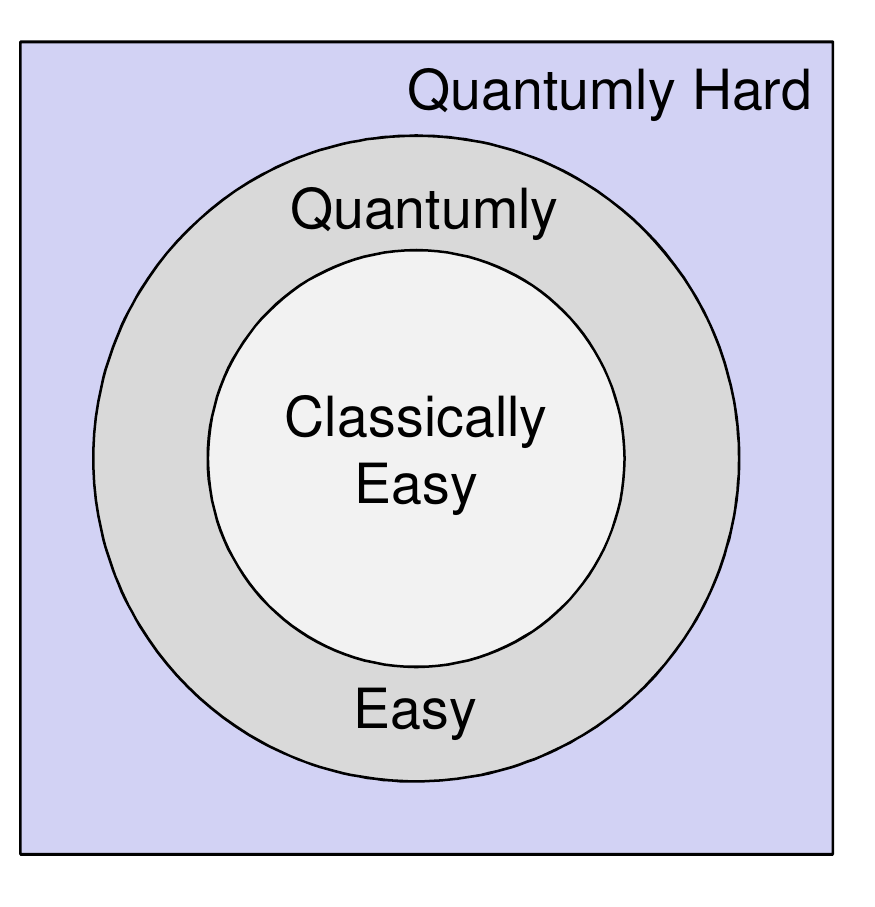} }}
\caption{(a) Hilbert space is vast, but the quantum states that can be prepared with reasonable resources occupy only a small part of it. (b) We believe that quantum computers can solve some problems that are hard for classical computers, but even quantum computers have limitations.} \label{fig:hilbert-hardness}
\end{figure}

The issue is subtle. Yes, the Hilbert space of a large quantum system is vast, because the classical description of a typical pure quantum state is enormously long. But we don't really care about typical quantum states, because preparing them is completely infeasible \cite{poulin11}. The only quantum states that are physically relevant are those that can be prepared with reasonable (quantum) resources, which are confined to an exponentially small portion of the full Hilbert space (Fig.~\ref{fig:hilbert-hardness}a). Only these can arise in Nature, and only these will ever be within the reach of the quantum engineers as technology advances. 

Mathematically, we may model the feasible quantum states this way: Imagine we have $n$ qubits (two-level quantum systems) which are initially in an uncorrelated product state. Then we perform a quantum circuit, a sequence of unitary operations (``quantum gates'') acting on pairs of qubits, where the total number of quantum gates is ``reasonable,'' let us say growing no faster than polynomially with $n$. Equivalently, we may say that a state is feasible if it can be constructed, starting with a product state, by evolving with a local Hamiltonian for a reasonable time. Likewise, we say a measurement is feasible if it can be constructed as a quantum circuit of size polynomial in $n$, followed by single-qubit measurements. 

These quantumly feasible states and measurements are plausibly allowed by Nature. Though far from ``typical,'' they may nevertheless be hard to simulate classically. That is why quantum computing is exciting and potentially powerful. 

\section{Separating classical from quantum}
The best evidence for such a separation between quantum and classical complexity comes from quantum algorithms that perform tasks going beyond what we know how to do with classical digital computers (Fig.~\ref{fig:hilbert-hardness}b). The most famous examples are Shor's algorithms for finding the prime factors of integers and evaluating discrete logarithms \cite{shor94}, which are based on using a fast quantum Fourier transform to probe the period of a function. 

There are other such ``superpolynomial'' speedups known, in which the time required to solve a problem scales polynomially with the input size when a quantum computer is used, but faster than polynomially when a classical computer is used. For example, by efficiently simulating topological quantum field theory using a quantum computer, we can evaluate approximately certain topological invariants of links and 3-manifolds ({\em e.g.}, the Jones polynomial \cite{freedman00,aharonov09} or Turaev-Viro invariant \cite{alagic10}). In fact, approximate evaluation of such topological invariants is a {\em BQP-hard} problem, meaning that any problem that a quantum computer can solve efficiently can be reduced to an instance of the problem of additively approximating the Jones polynomial of a link. 

A superpolynomial speedup is also achieved by a quantum algorithm for computing properties of solutions to systems of linear equations \cite{harrow09}. For example, if $A$ is an $N\times N$ Hermitian matrix, and $x$ solves $Ax=b$ where $x$ and $b$ are $N$-component vectors, then a quantum algorithm can estimate $x^\dagger M x$ in a time scaling like a power of $\log N$, provided $|b\rangle$ is an efficiently preparable quantum state, $A$ is sparse, and $M$ is an efficiently measurable operator. This problem, too, is BQP-hard. 

Someday, we hope to probe quantum physics in a previously unexplored regime by running fast quantum algorithms on quantum computers. For this purpose, it is convenient that the problems with superpolynomial speedups include some problems (like factoring) in the class NP, where the solution can be checked efficiently with a classical computer. Running the factoring algorithm, and checking it classically, we will be able to test whether Nature admits quantum processes going beyond what can be classically simulated. (However, this test is not airtight, because we have no proof that factoring is really classically hard.)

While quantum algorithms achieving superpolynomial speedups relative to classical algorithms are relatively rare, those achieving less spectacular polynomial speedups are more common. For example, a quantum computer can perform exhaustive search for a solution to a constraint satisfaction problem in a time scaling like the square root of the classical time \cite{grover96}, essentially because in quantum theory a probability is the square of an amplitude. By simulating a quantum walk on a graph, a quantum computer can also speed up the evaluation of a Boolean formula \cite{farhi07}, and hence determine, for example, whether a two-player game has a winning strategy. But again the speedup is merely polynomial. 

It seems that superpolynomial speedups are possible only for problems with special structure well matched to the power of a quantum computer. We do not expect superpolynomial speedups for the worst-case instances of problems in the NP class, such as 3-SAT or the Traveling Salesman Problem. For such problems with no obvious structure, we might not be able to do better than quadratically speeding up exhaustive search for a solution \cite{bernstein96}. 

But problems outside the class NP are also potentially of interest. Indeed, the ``natural'' application for a quantum computer is simulating evolution governed by a local Hamiltonian, preceded by the preparation of a reasonable state and followed by measurement of a reasonable observable \cite{feynman82}. In such cases the findings of the quantum computer might not be easy to check with a classical computer; instead one quantum computer could be checked by another, or by doing an experiment (which is almost the same thing).

As we strive toward the goal of quantum supremacy, it will be useful to gain a deeper understanding of two questions: (1) What quantum tasks are feasible? (2) What quantum tasks are hard to simulate classically? Conceivably, it will turn out that the extravagant exponential resources seemingly required for the classical description and simulation of generic quantum states are illusory; perhaps the quantum states realized in Nature really do admit succinct classical descriptions, either because the laws of physics governing complex quantum systems are different than we currently expect, or because there are clever ways to simulate the quantum world classically that have somehow eluded us so far. 

\section{Easiness and hardness}

Though we have sound reasons for believing that general quantum computations are hard to simulate classically, in some special cases the simulation is known to be easy. Such examples provide guidance as we seek a path toward quantum supremacy.

Suppose for example, that $n$ qubits are arranged in a line, and consider a quantum circuit such that, for any way of cutting the line into two segments, the number of gates that cross the cut is modest, only logarithmic in $n$. Then, if the initial state is a pure product state, the quantum state has a succinct classical description at all times, and the classical simulation of the quantum computation can be done efficiently \cite{vidal03,jozsa06}. The quantum computation does not achieve a super-classical task, because the quantum state becomes only slightly entangled.

Correspondingly, if you receive multiple copies of an $n$-qubit state that is only slightly entangled, you would be able to identify the state with a feasible number of measurements. In general, quantum state tomography is hard --- Hilbert space is so large that a number of measurements exponential in $n$ would be required to determine a typical $n$-qubit state. But for a slightly entangled state, a number of measurements linear in $n$ suffices \cite{cramer11}. We can perform tomography on segments of constant size, then do an efficient classical computation to determine how the pieces are stitched together. 

Gaussian quantum dynamics is also easy to simulate \cite{bartlett03}. Consider an interferometer assembled from linear optical elements, which can be described by a Hamiltonian quadratic in bosonic creation and annihilation operators. Suppose that a Gaussian initial state (a coherent state, for example) enters the input ports, and that we measure quadrature amplitudes at the output ports. Then the state has a succinct description at all times and can be simulated classically. But if we introduce some optical nonlinearity, or single photon sources together with adaptive photon counting measurements, then this system has the full power of a universal quantum computer, and presumably it cannot be simulated classically \cite{knill01,gottesman01}. Here ``adaptive'' means that subsequent operations can be conditioned on the outcomes of earlier measurements.

Free fermions are likewise easy to simulate classically, and, in contrast to free bosons, adaptive measurements of the fermion mode numbers do not add computational power \cite{valiant01,terhal02}. But if we add four-fermion operators to the Hamiltonian, or if we allow nondestructive measurements of four-fermion operators, then universal quantum computation is achievable \cite{bravyi00}.

\begin{figure}
\begin{center}
\includegraphics[width=0.8\textwidth]{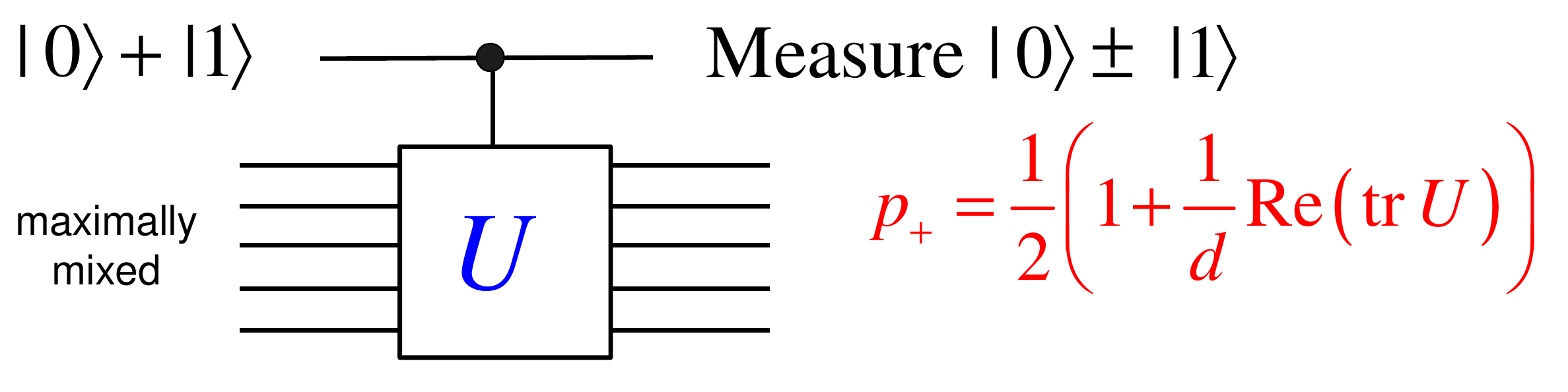}
\end{center}
\caption{\label{fig:trace} The trace of a large matrix can be computed in the ``one-clean-qubit'' model of quantum computation, for which the input is one pure qubit and many maximally mixed qubits.}
\end{figure}

\begin{figure}
\centering
\mbox{\subfigure[]{\includegraphics[width=0.40\textwidth]{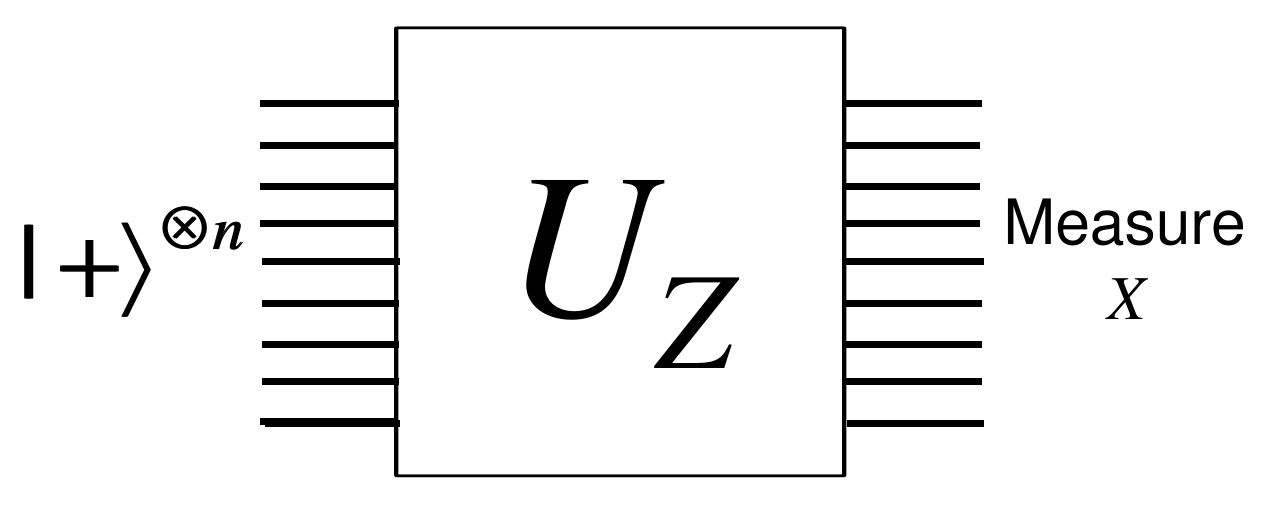}}
\hspace*{1in}
\subfigure[]{\includegraphics[width=.40\textwidth]{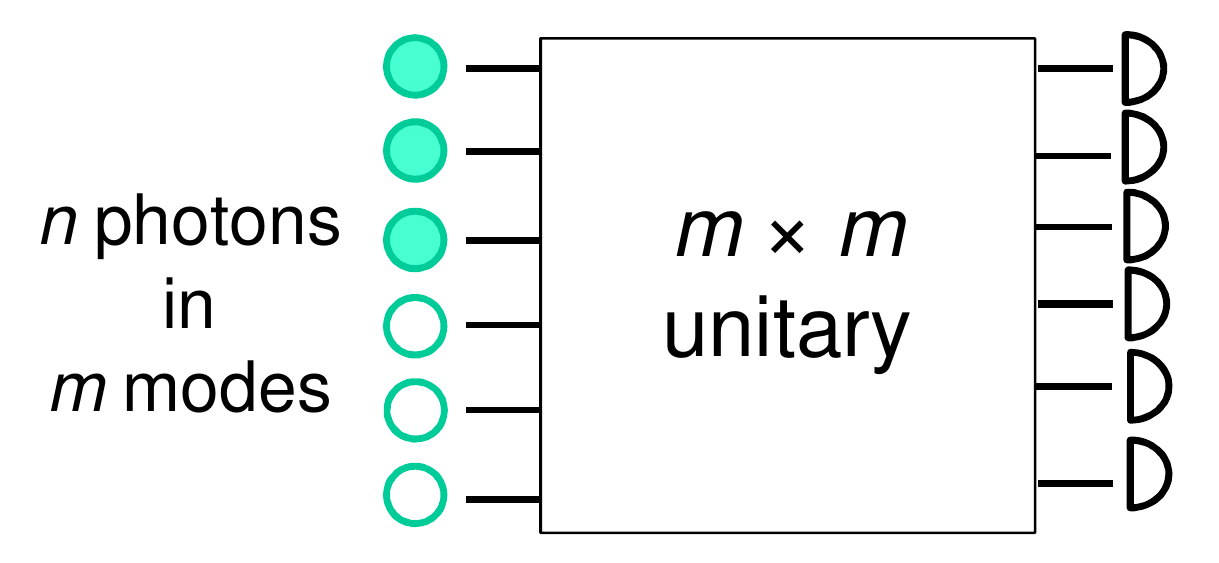} }}
\caption{Two quantum systems that may be hard to simulate classically. (a) A quantum circuit with commuting gates. (b) Nonadaptive linear optics with photon sources and photon detectors. } \label{fig:hard-to-simulate}
\end{figure}

Some computational models, though apparently weaker than the full blown quantum circuit model, nevertheless seem to have surprising power. One intriguing case is the ``one-clean-qubit model'', in which the input to the computation is one qubit in a pure state and many qubits in a maximally mixed state \cite{knill98}; see Fig.~\ref{fig:trace}. The study of this model was motivated initially by the nuclear-magnetic-resonance approach to quantum computing, where the initial state may be highly mixed \cite{gershenfeld97,cory97}. The one-clean-qubit quantum computer can evaluate the trace of an exponentially large unitary operator if the operator can be realized by an efficient quantum circuit. This capability can be exploited to approximate the Jones polynomial of the trace closure of a braid \cite{shor08} or the Turaev-Viro invariant of a three-dimensional mapping torus \cite{jordan09}, problems for which no efficient classical algorithms are known; in fact these problems are complete for the one-clean-qubit class. 

Another provocative example is the ``instantaneous quantum computing'' model \cite{bremner10}. Here all the gates executed by our quantum computer are mutually commuting, simultaneously diagonal in the standard $Z$ basis. In addition we can prepare single qubits in eigenstates of the conjugate operator $X$, and measure qubits in the $X$ basis; see Fig.~\ref{fig:hard-to-simulate}a. (Because all the gates commute, in principle they can be executed simultaneously.) It is not obvious how to simulate this simple quantum circuit classically, and there is evidence from complexity theory that the simulation is actually hard \cite{bremner10}. Even though the model does not seem to have the full power of universal quantum (or even classical) computing, nevertheless it may in a sense perform a super-classical task. 

Yet another tantalizing case is linear optics accompanied by photon sources and photon detectors, but now without adaptive measurements; see Fig. \ref{fig:hard-to-simulate}b. Suppose we have $m$ optical modes, where initially $n< m$ are occupied by single photons and the rest are empty. A linear optics array mixes the $m$ modes, and then a measurement is performed to see which of the output modes are occupied. Though this system is not a universal quantum computer, we do not know how to simulate it classically, and there is evidence from complexity theory that the simulation is hard \cite{aaronson10}. 

Such examples illustrate that there may be easier ways to achieve quantum supremacy than by operating a general purpose quantum computer. Admittedly, though, this linear optics experiment is still not at all easy --- to reach the regime where digital simulation is currently infeasible one should detect a coincidence of about 30 photons, whose paths through the interferometer can interfere. Furthermore, it is not clear how the hardness of simulating this system classically would be affected by including realistic noise sources, such as photon loss. 

\section{Local Hamiltonians}
An important task that a quantum computer can perform efficiently is simulating the dynamics of a quantum system governed by a local Hamiltonian $H$ \cite{lloyd96}. By ``local'' I do not necessarily mean {\em geometrically} local in some spatial dimension; instead, I mean that the Hilbert space has a decomposition into qubits (or other small systems), and $H$ can be expressed as a sum of terms, each of which acts on a constant number of qubits (independent of the system size). More generally, the simulation is feasible if $H$ is a sparse matrix \cite{aharonov03}.

\begin{figure}
\begin{center}
\includegraphics[width=0.7\textwidth]{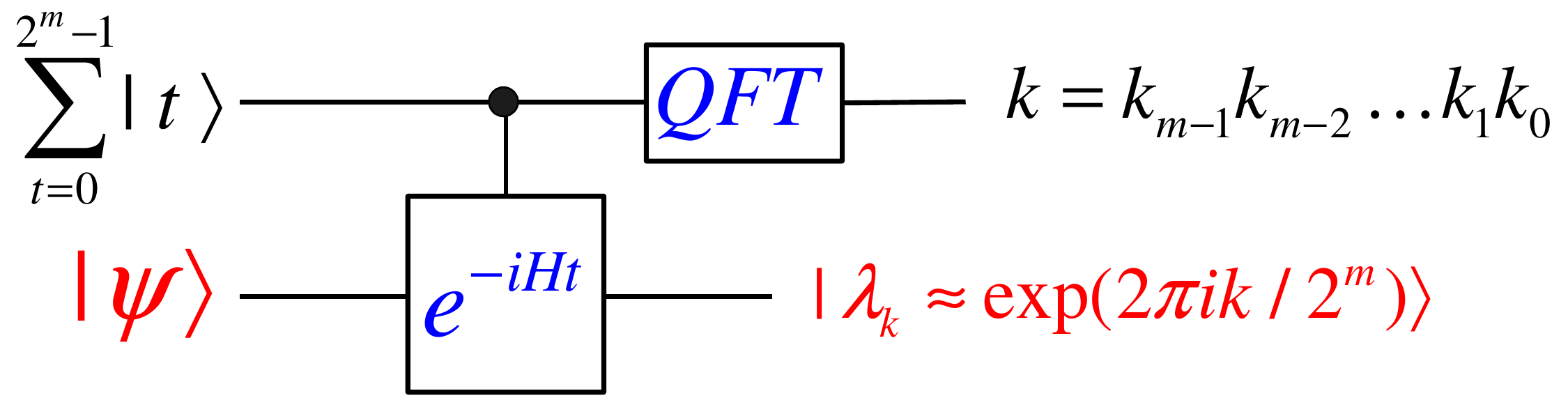}
\end{center}
\caption{\label{fig:phase-estimation} The energy of a system governed by a local Hamiltonian can be measured efficiently by a quantum computer, using a procedure called ``phase estimation.''}
\end{figure}

This capability can be exploited to measure the energy of the system, as in Fig.~\ref{fig:phase-estimation}. The quantum circuit shown evolves an initial state $|\psi\rangle$ for a time $t$ stored in an auxiliary register, then performs a quantum Fourier transform and reads out the register to sample from the frequency spectrum of the operator $e^{-iHt}$, a procedure called {\em phase estimation} \cite{kitaev95}.  The accuracy of the measured eigenvalue, in accord with the energy-time uncertainty relation, is inversely proportional to the maximal evolution time; hence, for an $n$-qubit system, accuracy scaling like an inverse polynomial in $n$ can be achieved by a quantum circuit with size polynomial in $n$. 

If the initial state $|\psi\rangle$ has an overlap with the ground state of $H$ which is not smaller than inverse polynomial in $n$, it follows that we can measure the ground-state energy to inverse polynomial accuracy in polynomial time using a quantum computer. This algorithm has noteworthy applications; for example, a quantum computer can compute the ground-state energy of a large molecule \cite{aspuru05}.

But there is a catch --- preparing an initial state that overlaps substantially with the ground state could be very hard in some cases. This is already true classically; finding the ground state of a classical spin glass is NP-hard, as hard as any problem whose solution can be checked efficiently by a classical computer. Finding the ground state for a quantum system with a local Hamiltonian seems to be even harder; it is QMA-hard \cite{kitaev02}, as hard as any problem whose solution can be checked efficiently by a quantum computer, and we expect that QMA is a larger class than NP. Surprisingly, computing the ground-state energy seems to be a hard problem for a quantum computer even for the case of a geometrically local translationally-invariant quantum system in one dimension \cite{gottesman-irani09}.

A general procedure for preparing ground states is adiabatic evolution. We can prepare a state having sizable overlap with the ground state of $H$ by starting with the easily prepared ground state of a simpler Hamiltonian $H(0)$, then slowly deforming the Hamiltonian along a path $H(s)$ connecting $H(0)$ to $H(1)=H$. This procedure succeeds in polynomial time provided the energy gap $\Delta(s)$ between the ground and first excited states of $H(s)$ is no smaller than inverse polynomial in $n$ for all $s\in[0,1]$ along the path. For problem instances that are quantumly hard, then, the gap becomes superpolynomially small somewhere along the path \cite{farhi00}.

Though the general problem is quantumly hard, we may surmise that there are many local quantum systems for which computing the ground-state energy is quantumly easy yet classically hard. Furthermore, a quantum computer may be able to simulate the evolution of excited states in cases where the simulation is classically hard, such as chemical reactions \cite{aspuru08} or the scattering of particles described by quantum field theory \cite{jordan11}. Even in the case of quantum gravity, evolution may be governed by a local Hamiltonian, and therefore admit efficient simulation by a quantum computer. 

\section{Quantum error correction}
\label{sec:qec}

Classical digital computers exist, and have had a transformative impact on our lives. Large-scale quantum computers do not yet exist. Why not?

Building reliable quantum hardware is challenging because of the difficulty of controlling quantum systems accurately. Small errors in quantum gates accumulate in a large circuit, eventually leading to large errors that foil the computation. Furthermore, qubits in a quantum computer inevitably interact with their surroundings; decoherence arising from unwanted correlations with the environment is harmless in a classical computer (and can even be helpful, by introducing friction which impedes accidental bit flips), but decoherence in a quantum computer can irreparably damage the delicate superposition states processed by the machine.

Quantum information might be better protected against noise by using a quantum error-correcting code, in which ``logical'' information is encoded redundantly in a block of many physical qubits \cite{shor95,steane95}. Quantum error correction is in some ways much like classical error correction, but more difficult, because while a classical code need only protect against bit flips, a quantum code must protect against both bits flips and phase errors. 

Suppose for example, that we want to encode a single logical qubit, with orthonormal basis states denoted $|0\rangle$ and $|1\rangle$, which is protected against all the errors spanned by a set $\{E_a\}$. For the distinguishability of the basis states to be maintained even when errors occur, we require
\be\label{eq:bit-perp}
E_a|0\rangle \perp E_b|1\rangle,
\ee
where $E_a,E_b$ are any two elements of the error basis. This condition by itself would suffice for reliable storage of a classical bit.

But for storage of a qubit we also require protection against phase errors, which occur when information about whether the state is $|0\rangle$ or $|1\rangle$ leaks to the environment; equivalently, distinguishability should be maintained for the dual basis states $|0\rangle \pm |1\rangle$:
\be\label{eq:phase-perp}
E_a\left(|0\rangle +|1\rangle\right)\perp E_b\left(|0\rangle -|1\rangle\right),
\ee
where $E_a,E_b$ are any two errors. In fact, the two distinguishability conditions Eq.~(\ref{eq:bit-perp}) and (\ref{eq:phase-perp}) suffice to ensure the existence of a recovery map that corrects any error spanned by $\{E_a\}$ acting on any linear combination of $|0\rangle$ and $|1\rangle$ \cite{knill-laflamme97}.

Together, Eq.~(\ref{eq:bit-perp}) and (\ref{eq:phase-perp}) imply
\be\label{eq:bit-same}
\langle 0| E_a^\dagger E_b |0\rangle = \langle 1| E_a^\dagger E_b |1\rangle;
\ee
no measurement of any operator in the set $\{E_a^\dagger E_b\}$ can distinguish the two basis states of the logical qubit. Typically, because we expect noise acting collectively on many qubits at once to be highly suppressed, we are satisfied to correct {\em low-weight} errors, those that act nontrivially on a sufficiently small fraction of all the qubits in the code block. Then Eq.~(\ref{eq:bit-same}) says that all the states of the logical qubit look the same when we examine a small subsystem of the code block. These states are highly entangled, like the hundred-page book that reveals no information when we read the individual pages. 

\begin{figure}
\begin{center}
\includegraphics[width=0.7\textwidth]{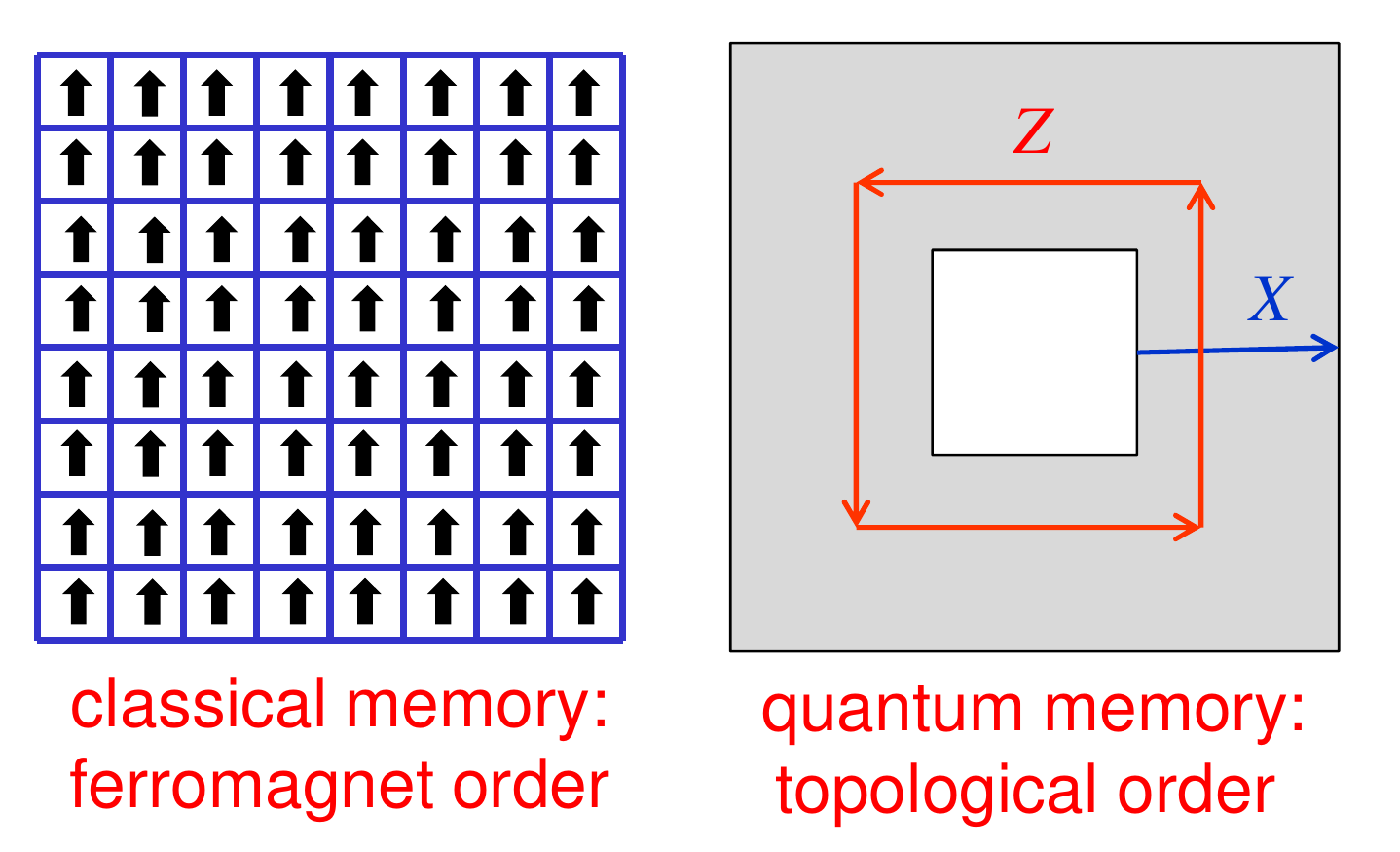}
\end{center}
\caption{\label{fig:memory} A prototypical classical memory is a ferromagnet, and a prototypical quantum memory is a topologically ordered medium.}
\end{figure}

It is useful to formulate the distinction between classical and quantum error correction in more physical terms (see Fig.~\ref{fig:memory}). The prototype for a protected classical memory is a ferromagnet, where a single bit is encoded according to whether most of the spins are up or down. The encoded bit can be read out by performing local measurements on all spins, and then executing a majority vote to protect against errors that flip a minority of the spins. Errors in the memory create domain walls where neighboring spins misalign, and a logical error occurs when a domain wall sweeps across the sample, inducing a global operation acting on many spins. The memory is robust at a sufficiently small nonzero temperature because the energy cost of a large droplet of flipped spins is large. This memory is a particularly simple physically motivated example of a classical error-correcting code; there are more sophisticated examples. 

The prototype for a protected quantum memory is a medium in two dimensions with $Z_2$ topological order \cite{kitaev03}. We may consider a planar sample with a large hole in the middle. In contrast to the ferromagnet, errors in the medium create pointlike quasiparticles (``anyons'') rather than domains walls. There are two types of anyons (which we may regard as ``electric'' and ``magnetic'' excitations), having $Z_2$ Aharonov-Bohm interactions with one another. The space of quantum states with no particles present is two-dimensional --- this space is the encoded qubit. Logical errors can be induced by the transport of particles; a logical $X$ acts on the encoded qubit if an electric particle travels between the inner and outer boundaries of the sample, and a logical $Z$ error acts if a magnetic particle travels around the hole. Correspondingly, we read out the logical qubit in the $X$ basis by measuring a nonlocal string-like operator which connects the inner and outer boundaries, simulating the propagation of an electric particle, while we read it out in the $Z$ basis by measuring a string operator that encloses the hole in the sample, simulating the propagation of a magnetic particle.

The system is protected by a nonzero energy gap, the energy cost of creating a pair of particles. Hence the storage time is long if the temperature is small compared to the gap, but unlike the case of a two-dimensional ferromagnet the storage time does not improve as the system size increases. However, if we monitor the particles as they diffuse through the sample, then a logical error occurs only if particles propagate across the sample without being noticed, an event which {\em does} become increasingly unlikely as the system size grows \cite{dennis02}. 

A topologically ordered medium on a topologically nontrivial surface is a special type of quantum error-correcting code, one that can be realized as the ground state of a system with a geometrically local Hamiltonian; in this respect its status is similar to that of the ferromagnet in classical coding theory. The locality of the Hamiltonian has advantages. For one, we might be able to realize a relatively robust quantum memory described by a Hamiltonian in the universality class of the code. From a more abstract viewpoint, we can collect information about the errors in the code block by making localized measurements, {\em e.g.}, by identifying domain walls in the ferromagnet or quasiparticle excitations (anyons) in the topologically ordered medium.

\section{Scalable quantum computing}
The theory of quantum error correction establishes that quantum computing is ``scalable'' in principle. This means that, if the noise strength is below a critical value (the ``accuracy threshold''), then we can simulate an ideal quantum circuit accurately using a circuit of noisy gates, with a reasonable overhead cost in additional gates and additional qubits \cite{aharonov97,kitaev97,laflamme97,aliferis06,reichardt06}. The numerical value of the threshold, and the overhead cost, depend on the fault-tolerant scheme used and on how we model the noise.

Engineering considerations favor a two-dimensional layout with short-range interactions among the qubits, for which the computation can be protected against noise by using a topological code like the one described in Sec.~\ref{sec:qec}. A topological medium can be simulated using any convenient type of quantum hardware, with the physical qubits carried by, for example, trapped ions, electron spins in quantum dots, or superconducting circuits. To encode many logical qubits, the simulated medium has many holes, and logical errors are suppressed by ensuring that the holes are sufficiently large and distantly separated from one another.  A complete set of universal quantum gates can be executed on the encoded qubits; hence arbitrary quantum circuits can be simulated efficiently and reliably \cite{dennis02,raussendorf07}.


There are many challenges to making large-scale fault-tolerant quantum computing practical, including serious systems engineering issues. There are also issues of principle to consider, such as, what is required for a fault-tolerant scheme to be scalable, and what conditions must be satisfied by the noise model? One essential requirement is some form of cooling, to extract the entropy introduced by noise \cite{aharonov96}. Parallel operations are also necessary, so noise can be controlled in different parts of the computer simultaneously.

It is natural to describe noise using a Hamiltonian that includes a coupling between the system and its unobserved environment, and proofs of scalability require the noise to be suitably local. For example, we may write the noise Hamiltonian as a sum of terms, each acting on just a few of the physical qubits in the quantum computer, but possibly acting on the environment in a complicated way. Then the proof of scalability applies if each such term in the noise Hamiltonian has a sufficiently small norm \cite{terhal05,aliferis06,aharonov06}. If the noise Hamiltonian includes terms that act on $k >> 1$ qubits in the quantum computer (and in some complicated way on the environment), the proof of scalability works if these terms decay exponentially with $k$, and also decay rapidly enough as the qubits separate in space. A drawback of such scalability criteria is that the condition on the noise is not expressed in terms of directly measurable properties; an advantage is that the state and dynamics of the environment need not be specified.

Alternatively, we may suppose that the environment is described by a Gaussian free field, so the noise can be completely characterized by its two-point correlation function. Then the proof of scalability goes through if the noise is sufficiently weak, with correlations decaying sufficiently rapidly in both time and space \cite{ng09}. This criterion has the advantage that it is expressed in terms of measurable quantities, but it applies only for if the initial state and the dynamics of the environment obey suitable restrictions. 

Thus quantum error correction works in principle for noise that is sufficiently weak and not too strongly correlated, but may fail if the noise acts collectively on many qubits at once. As quantum hardware continues to advance, it will be important to see whether the noise in actual devices has adequately weak correlations, keeping in mind that there are possible ways to suppress correlations, for example by using dynamical decoupling sequences \cite{viola99}.

\section{Topological quantum computing}
To a theorist, a particularly appealing and elegant way to achieve fault-tolerant quantum computing is by using the exotic statistics of nonabelian anyons \cite{kitaev03,ogburn98,freedman02}. Quantum information, stored in the exponentially large fusion Hilbert space of $n$ anyons, is well protected if the temperature is low compared to the energy gap (to prevent unwanted thermal production of anyon pairs) and if the anyons are kept far apart from one another (to prevent unwanted nontopological interactions due to quantum tunneling). Robust information processing can be achieved by exchanging the particles, exploiting their exotic quantum statistics, and information can be read out by measuring charges of anyon pairs (for example, using an interferometer \cite{halperin06,bonderson06}). 

An early proposal for achieving quantum computing with anyons was based on fractional quantum Hall states \cite{dassarma05,nayak08}; more recent 
proposals exploit exotic properties of topological superconductors and topological insulators \cite{kitaev01,fu08,sau10,alicea10,lutchyn10,oreg10,alicea11}. In most such proposals, the anyon braiding by itself is not sufficient for universal quantum computing, but can be supplemented by unprotected (and possibly quite noisy) nontopological operations to realize a universal gate set \cite{bravyi05}. Indeed, in some cases \cite{alicea11} braiding of anyons can be modeled faithfully by a time-dependent free-fermion Hamiltonian; therefore, the nonuniversality of braiding operations follows from the observation that free-fermion systems can be simulated classically, together with the presumption that efficient classical simulations of general quantum circuits are impossible.

Since the error rate is suppressed by the energy gap for anyon pair creation, and does not improve as the system size increases, we may anticipate that for very large-scale applications topological quantum computing will need to be supplemented by ``standard'' methods of quantum error correction. However, if topological protection enforces a very low gate error rate, the overhead cost of using quantum error-correcting codes may be relatively modest. 

Classical information in a ferromagnet is protected ``passively,'' because memory errors occur only when the system surmounts an energy barrier whose height increases sharply with system size. Could there be topologically ordered quantum systems that likewise store quantum information passively, providing a mechanism for a ``self-correcting'' quantum memory? \cite{bacon06} Models realizing this vision are known in four spatial dimensions \cite{dennis02,alicki10,chesi10}. A recently discovered three-dimensional quantum model has a barrier height increasing logarithmically with system size, but for this model the storage time is bounded above, and declines once the system grows beyond an optimal size \cite{haah11,bravyi11}.

\section{Quantum computing vs. quantum simulation}

One of the most important applications for quantum computing will be simulating highly entangled matter such as quantum antiferromagnets, exotic superconductors, complex biomolecules, bulk nuclear matter, and spacetime near singularities. A general purpose quantum computer could function as a ``digital'' quantum simulator, in contrast to ``analog'' quantum simulators based on customizable systems of (for example) ultracold atoms or molecules. The goal of either digital or analog quantum simulation should be achieving quantum supremacy, {\em i.e.}, learning about quantum phenomena that cannot be accurately simulated using classical systems. In particular, we hope to discover new and previously unsuspected phenomena, rather than just validate or refute predictions made by theorists. 

A universal quantum computer will be highly adaptable, capable of simulating efficiently any reasonable physical system, while analog quantum simulators have intrinsic limitations. In particular, it is not clear to what degree the classical hardness hinges on the accuracy of the simulation, and present day quantum simulators, unlike the universal quantum computers of the future, are not fault tolerant. On the other hand, analog quantum simulators may be able to probe, at least qualitatively, exotic quantum phenomena that are sufficiently robust and universal as to be studied without tuning the Hamiltonian precisely. Furthermore, since the characteristic imperfections in analog quantum simulations vary from one experimental platform to another, obtaining compatible results using distinct simulation methods will boost confidence in the results. 


\section{Conclusions and questions}

I have emphasized the goal of quantum supremacy (super-classical behavior of controllable quantum systems) as the driving force behind the quest for a quantum computer, and the idea of quantum error correction as the basis for our hope that scalable quantum computing will be achievable. 
To focus the talk, I have neglected other deeply engaging themes of quantum information science, such as quantum cryptography and the capacities of quantum channels. Also, I have not discussed the impressive progress in building quantum hardware, a topic covered by other rapporteurs. I'll conclude by raising a few questions posed or suggested in the preceding sections.

Regarding quantum supremacy, might we already have persuasive evidence that Nature performs tasks going beyond what can be simulated efficiently by classical computers? For example, there are many mathematical questions we cannot answer concerning strongly correlated materials and complex molecules, yet Nature provides answers; have we failed because these problems are intrinsically hard classically, or because of our lack of cleverness so far?

Is quantum simulation (e.g. with cold atoms and molecules) a feasible path to quantum supremacy? Or will the difficulty of controlling these systems precisely prevent us from performing super-classical tasks? 

How can we best achieve quantum supremacy with the relatively small systems that may be experimentally accessible fairly soon, systems with of order 100 qubits? In contemplating this issue we should keep in mind that such systems may be too small to allow full blown quantum error correction, but also on the other hand that a super-classical device need not be capable of general purpose quantum computing.



Regarding quantum error correction, what near-term experiments studying noise in quantum hardware will strengthen the case that scalable fault-tolerant quantum computing is feasible? What pitfalls might thwart progress as the number of physical qubits scales up? 

Do the observed properties of topologically ordered media such as fractional quantum Hall systems and topological superconductors already provide strong evidence that highly robust quantum error-correcting codes are physically realizable? How much more persuasive will this evidence become if and when the exotic statistics of nonabelian anyons can be confirmed directly?

Which is a more promising path toward scalable quantum computing: topological quantum computing with nonabelian anyons, or fault-tolerance based on standard qubits and quantum error-correcting codes? Will the distinction between these two approaches fade as hardware advances? 

Can a quantum memory, like a classical one, be self-correcting, with storage time increasing as the system grows? Can quantum information protected by self-correcting systems be processed efficiently and reliably?



How might quantum computers change the world? Predictions are never easy, but it would be especially presumptuous to believe that our limited classical minds can divine the future course of quantum information science. Attaining quantum supremacy and exploring its consequences will be among the great challenges facing 21st century science, and our imaginations are poorly equipped to envision the scientific rewards of manipulating highly entangled quantum states, or the potential benefits of advanced quantum technologies. As we rise to the call of the entanglement frontier, we should expect the unexpected.

\section*{Acknowledgments}
I am grateful to the organizers for the opportunity to attend this exciting meeting. This work was supported in part by NSF grant PHY-0803371, DOE grant DE-FG03-92-ER40701, and NSA/ARO grant W911NF-09-1-0442.


\end{document}